\documentstyle[preprint,tighten,aps,epsfig]{revtex}

\def\lapprox{\mathrel{\mathop  {\hbox{\lower0.5ex\hbox{$\sim$}
\kern-0.8em\lower-0.7ex\hbox{$<$}}}}}  

\def\gapprox{\mathrel{\mathop  {\hbox{\lower0.5ex\hbox{$\sim$}
\kern-0.8em\lower-0.7ex\hbox{$>$}}}}}

\begin{document}

\draft

\preprint{\vbox{\noindent{}\hfill INFNFE-10-02}}

\title{Fusion rate enhancement due to energy spread of colliding nuclei}

\author{G.~Fiorentini$^{(1,2)}$, C. Rolfs$^{(3)}$, F.L. Villante$^{(1,2)}$ 
and B. Ricci $^{(1,2)}$}

\address{
$^{(1)}$Dipartimento di Fisica dell'Universit\`a di Ferrara,I-44100
Ferrara, Italy\\
$^{(2)}$Istituto Nazionale di Fisica Nucleare, Sezione di Ferrara, 
I-44100 Ferrara,Italy \\
$^{(3)}$ Institut f\"{u}r Experimentalphysik III, Ruhr-Universit\"{a}t Bochum, Germany}

\date{\today}

\maketitle

\begin{abstract}
Experimental results for sub-barrier nuclear fusion reactions show
cross section enhancements with respect to bare nuclei which are
generally larger than those expected according to electron screening
 calculations. We point out that energy spread of target or projectile nuclei
is a mechanism which generally provides fusion enhancement. We present a
general formula for calculating the enhancement factor and we provide
quantitative estimate for effects due to thermal motion, vibrations inside
atomic, molecular or crystal system, and due to finite beam energy width.
All these effects are marginal at the energies which are presently           
measurable, however they have to be considered  in                    
future experiments at still lower energies.
This study allows to exclude several effects as possible    
explanation of the observed anomalous fusion enhancements,         
which remain a mistery.
\end{abstract}

\section{Introduction}
\label{intro}
The chemical elements were created by nuclear fusion reactions in 
the hot interiors of remote and long-vanished stars over  many
billions of years \cite{RolfsRodney}. Thus, nuclear reaction 
rates  are at the heart of nuclear astrophysics: they influence 
sensitively the nucleosynthesis of the elements in the earliest 
stages of the universe and in all the objects formed thereafter, 
and they control the associated energy generation, neutrino luminosity 
and evolution of stars. A good knowledge of their rates
 is essential for understanding this broad 
picture.

Nuclear reactions in static stellar burning phases
  occur at energies far below 
the Coulomb barrier. Due to the steep drop  of the cross section
$\sigma(E)$  at sub-barrier energies, it becomes increasingly 
difficult to measure it  as the energy $E$ is lowered.
Generally, stellar fusion rates are obtained by extrapolating 
laboratory data taken at energies significantly larger  than those 
relevant to stellar interiors. Obviously such an ``extrapolation into 
the unknown" can lead to considerable uncertainty. In the last 
twenty years a significant effort has been devoted to the 
experimental exploration of the lowest energies and new approaches 
have been developed so as to reduce the uncertainties in the 
extrapolations. In particular, the installation of an accelerator 
facility in the underground laboratory at LNGS \cite{Luna1} 
has allowed the  $\sigma (E)$ measurement of  
$^3{\rm He}(^3{\rm He},2{\rm p})^4{\rm He}$ down 
to its solar  Gamow peak, $E_0 \pm \Delta /2 = (21 \pm5)$ keV \cite{Luna2} so 
that for this reaction no extrapolation is needed anymore.

As experiments have moved well down into the  sub-barrier region, 
the screening effect of atomic electrons has become relevant 
\cite{Rolfs87,Bracci90,Rolfs2001}. With 
respect to the bare nuclei case, the Coulomb repulsion is 
diminished, the tunneling distance $R_{\rm t}$ is reduced, and the fusion 
probability, which depends exponentially on $R_{\rm t}$, is enhanced.
The electron effect on the reaction  can be seen as a transfer of 
energy $U$ (the screening potential energy) from the electronic to the 
translational degrees of freedom.  For each collision energy $E$, 
one has an effective energy  $E_{\rm eff}=E+U$ and a cross section 
enhancement:
\begin{equation}
\label{effe}
f = \sigma(E+U)/\sigma(E)
\end{equation}

The screening potential energy $U$ is easily estimated in two limiting 
cases \cite{Bracci90}:
 In the sudden limit, when the relative velocity $v_{rel}$ 
of the nuclei 
is larger than the typical electron velocity $v_0=e^2/ \hbar$:
the electron wave function during the nuclear collision  is 
frozen at the initial value  $\Psi_{\rm in}$ and the energy transferred 
from electrons to the  nuclei is thus
\begin{equation}
\label{Usu}
U_{\rm su} = \langle \Psi_{\rm in}  |Z_1  e^2/r_{1e} | \Psi_{\rm in} \rangle \quad , 
\end{equation}
where here and in the following 
the index 1 (2) denotes the projectile (target) nucleus and 
a sum over the electrons is understood.
In the adiabatic limit, i.e. when $v_{rel} \ll v_0$: 
electrons follow 
adiabatically the nuclear motion and  at any internuclear distance 
the electron wave function $\Psi_{\rm ad}$ corresponds to  an 
energy eigenstate calculated for fixed nuclei. As the nuclei approach 
distances smaller than each atomic radius , $\Psi_{\rm ad}$  tends 
to the united atom (i.e. with nuclear charge $Z=Z_1+Z_2$) limit,  
$\Psi_{\rm un}$. The kinetic energy gained by the colliding nuclei is 
thus
\begin{equation}
U_{\rm ad} = \epsilon_{\rm in} - \epsilon_{\rm un} \quad ,
\end{equation}
where $\epsilon_{\rm in}$ ($\epsilon_{\rm un}$) is the electron energy of the 
isolated (united) atom in the corresponding states.

We like to stress a few important features:

{\em i)} Screening potential energies, which are in the range 10--100 eV, are 
definitely smaller than the practical collision energies (1--100 
keV), nevertheless they can produce appreciable fusion 
enhancements due to the exponential dependence of the cross 
section.

{\em ii)} In the adiabatic limit the electron energy assumes the lowest 
value consistent with quantum mechanics. Due to energy 
conservation, the energy transfer to the nuclear motion is thus 
maximal in this case ($U<U_{\rm ad}$) and the observed cross section 
enhancement should not exceed that calculated by using the 
adiabatic potential:
\begin {equation}
\label{fad}
f \le f_{\rm ad} = \sigma(E+U_{\rm ad})/\sigma(E)
\end {equation}

{\em iii)} The enhancement factors which have been measured are generally 
larger than expected.  A summary of the available results is 
presented in Table \ref{Tabrolf}. The general trend is that the enhancement 
factors exceed the adiabatic limit. Recent measurement of $d(d,p)^3{\rm H}$
 with deuterium implanted in metals \cite{deuterium} have shown 
enhancements of the cross sections with respect to the bare nuclei 
 case by factors of order unity, whereas one expects a few percent
effect.
In other words, if one  derives   an ``experimental'' potential energy
$U_{\rm ex}$ from a fit of experimental data according to eq. (\ref{effe}),  
the resulting values significantly exceed the adiabatic limit 
$U_{\rm ad}$. In the case of deuterium implanted in  metals, 
values  as high as $U_{\rm ex} \simeq  700$ eV  have been found \cite{deuterium}, 
at least an order of 
magnitude larger than the expected atomic value $U_{ad}$.
Several theoretical investigations have resulted in a better 
understanding of small effects in low energy nuclear reactions, but have not 
provided an explanation of this puzzling picture.

{\em iv)} Dynamical calculations of electron screening for finite values of 
the relative velocity  show a smooth interpolation between the 
extreme adiabatic and sudden limits
\cite{BraccieFiorentinTAUP,shoppa}. In fact, one cannot 
exceed  the value obtained in the adiabatic approximation because 
the dynamical calculation includes atomic excitations which reduce  
the energy transferred  from the electronic binding to the 
relative motion.

{\em v)} The effects of vacuum polarization \cite{Scilla1994}, 
relativity, Bremmstrahlung 
and atomic polarization \cite{Balantekin1997} 
have been studied.  Vacuum polarization becomes  relevant when  
the minimal approach distance is close to the electron Compton 
wavelength but it has an anti-screening effect, corresponding 
to the fact that in QED the effective charge increases at short 
distances. All these effects cannot account for the anomalous 
enhancements.

Although one cannot exclude some experimental effect, e.g. a 
(systematic) overestimate of the stopping power, the  general 
trend is that most reactions exhibit an anomalous high enhanchement.
Phenomenologically, this corresponds to an unexplained collision 
energy increase in the range of 100 eV. 

Actually, the anomalous experimental values $U_{\rm ex}$ look too 
large to be related with atomic, molecular or crystal energies.  
Some other processes, involving the much smaller energies available 
in the target, should mimic the large experimental  values of U.
As an example, if the projectile approaches a  target nucleus 
which is moving against it with energy $E_2 \ll E$, the 
collision energy is increased by an amount:
\begin{equation} 
U  = \left ( \frac{4 m_1}{m_1+m_2} \,E\, E_2 \right )^{1/2}.
\label{E_gain}
\end{equation}
For $d+d$ reactions at (nominal) collision energy $E=10$~keV, a 
target energy $E_2=0.5$ eV is sufficient for producing $U=100$~eV.

Generally, one expects that opposite motions of the target  nuclei 
are equally possible. Even in this case, however, the effect is 
not washed out: due to the strong nonlinearity of the fusion 
cross section the reaction probability is much larger for those 
nuclei which are moving against the projectile.

In this spirit, we shall consider processes associated with the 
energy spread of the colliding nuclei. These processes generally 
lead to an enhancement of the fusion rate, for the reasons just 
outlined. 

In the next section we shall first consider the thermal motion of 
the target nuclei. For this  example, we 
shall derive an expression for the enhancement factor on physical 
grounds and then we shall outline the effects of an energy spread 
for the extraction of the astrophysical S-factor from experimental 
data.

The treatment is generalized in sect. \ref{general} and  in sect. \ref{appli}
 it is applied 
to study energy spreads due to  motion of the nuclei inside atoms, 
molecules and crystals. Beam energy width and straggling are also 
considered.

In summary, all the effects turn out to be too tiny to explain the observed anomalous 
enhancements. Nevertheless, they have  to be considered in 
analyzing the data, particularly in  future experiments at still 
lower energies.

\section{The effect of thermal motion of target nuclei}
\label{sec2}

In this  section we consider the effects of thermal motion of the 
target nuclei. We shall make several simplifications, in order to 
elucidate the main physical ingredients. In this way we shall 
derive a simple expression for the enhancement factor on physical 
grounds. 

Essentially, we shall concentrate on the exponential factor of the 
fusion cross section, neglecting the energy dependence of 
the pre-exponential factors, and we shall only consider the effect 
of the target motion in the direction of the incoming particle, 
neglecting the transverse motion. When these simplifications are 
removed the result is essentially confirmed:
see the more general treatment of sect.~\ref{general}. 

The fusion cross section at energies well below the Coulomb 
barrier is generally written as:
\begin {equation}
\label{sigmadiE}
\sigma = \frac{S(E)}{E} \, \exp \left (-\frac{V_0}{\sqrt{2E/\mu}} \right )
\end{equation}
where $E=\frac{1}{2} \mu v_{\rm rel}^2$ is the collision energy, 
$\mu=m_1 m_2/(m_1+m_2)$ is the reduced mass, $V_0=Z_1 Z_2 e^2/ \hbar$ 
and $S(E)$ is the astrophysical S-factor
\footnote{For convenience of the reader, we recall that 
$v_0=e^2/\hbar$ and thus $V_0=Z_1 Z_2 v_0$. }.
The cross section
 is more conveniently 
expressed in terms of the relative velocity of the colliding 
nuclei $v_{\rm rel}$:
\begin{equation}
\label{sigmadivrel}
\sigma(v_{\rm rel}) 
=\frac{2S(v_{\rm rel})}{\mu v_{\rm rel}^2} \, \exp \left (-\frac{V_0}{v_{\rm rel}} \right ) 
\end{equation}
At energies well below the Coulomb barrier, 
$ v_{\rm rel} \ll V_0 $, the 
main dependence is through the exponential factor, so we shall 
treat the pre-exponential term as a constant:
\begin{equation}
\sigma(v_{\rm rel}) \simeq  B \, \exp \left (-\frac{V_0}{ v_{\rm rel}} \right ) \quad .
\end{equation}

We consider  a projectile nucleus with fixed velocity $\mathbf{V}$ impinging 
against a target where the nuclei have a thermal distribution 
of velocity.
Since the target nucleus velocity $\mathbf{v}$
 is generally much smaller than $V=|\mathbf{V}|$,   one can 
expand $1/v_{\rm rel}= 1/|\mathbf{V}-\mathbf{v}|$ and retain 
the first non vanishing term:
\begin{equation}
\label{sigmaexpan}
\sigma \simeq  B \, 
\exp \left (-\frac{V_0}{V} - \frac{V_0 v_{\parallel}}{V^2} \right ) \quad ,
\end{equation}
where $v_\parallel$ is the target velocity projection over the $\mathbf{V}$-direction.

\begin{figure}
\centerline{\hbox{
\epsfig{figure=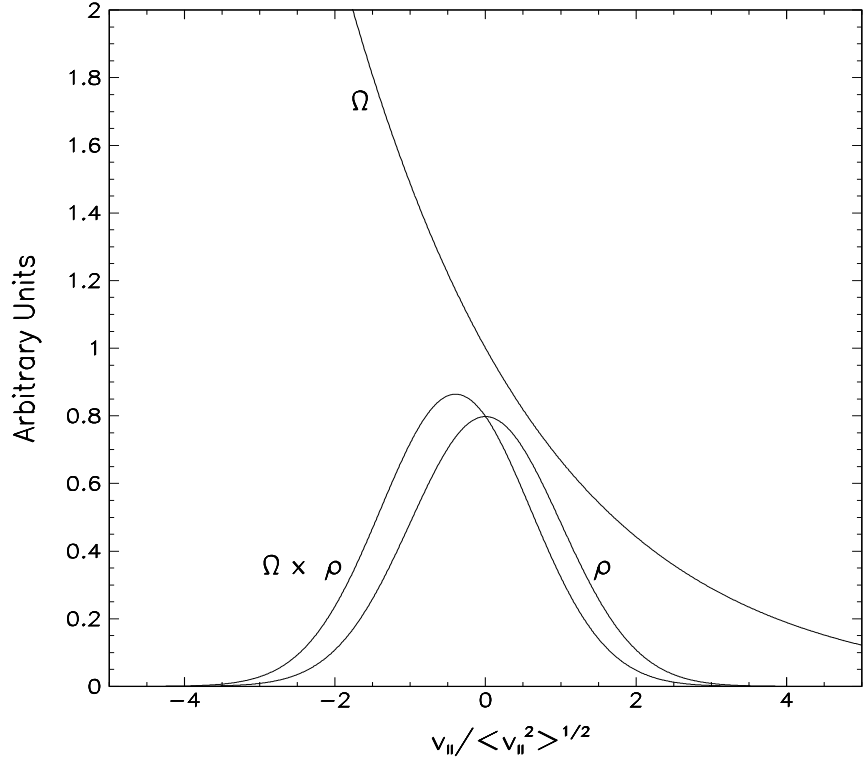,height=15cm,width=12cm}}}
\vskip-2cm
\caption[a]{A sketch of the contribution to the averaged cross section.
$\rho(v_\parallel)$ is defined in eq.(\ref{thermal}) 
and $\Omega=exp(-v_0 \langle v_\parallel \rangle /V^2)$.}
\label{figgamow}
\end{figure}

The enhancement factor with respect to the fixed target case, 
$f=\langle \sigma\rangle /\sigma(V)$, is thus calculated by averaging 
$\exp(-V_0 v_{\parallel}/V^2)$ over the $v_\parallel$ distribution:
\begin{equation}
\label{thermal}
\rho(v_\parallel) = \frac{1}{\sqrt{2\pi \langle v_\parallel ^2\rangle  } } \, 
\exp \left (-\frac{1}{2}\, \frac{v_\parallel^2}{\langle v_\parallel ^2\rangle } \right ),
\end{equation}
where $\langle v_{\parallel}^2\rangle =kT/m_2$.
The integral
\begin{equation}
\label{fthermaleq}
f =  \frac{1}{\sqrt{2\pi \langle v_\parallel ^2\rangle  } } \,
\int_{-\infty}^{+\infty}dv_\parallel \,  \exp \left (- \frac{V_0 v_\parallel}{ V^2} - 
\frac{v_\parallel^2}{2\langle v_\parallel ^2\rangle } \right )
\end{equation}
is easily evaluated by using a (saddle point) trick similar to 
that used by Gamow for evaluating stellar burning rates. The 
product of the Gaussian and the exponential functions (Fig.\ref{figgamow})
results in an (approximately) Gaussian with the same width, centered at 
$v_{\rm G}=-\langle v_\parallel ^2\rangle  V_0/V^2$, its  height  giving the enhancement factor:
\begin{equation}
\label{fthermal}
f = \exp\left (\frac{V_0^2 \langle v_\parallel^2\rangle }{2 V^4} \right ) \quad .
\end{equation}

Concerning this equation, which is the main result of the paper, 
several comments are needed:

{\em i)} Since the term in parenthesis in eq. (\ref{fthermal}) is positive, 
one has $f\geq 1$, 
i.e.  the energy spread always results in a cross section 
enhancement. One cannot ignore 
the target velocity distribution for the calculation of the 
reaction yield since  nuclei moving towards the projectile 
have a larger weight in the cross section.

{\em ii)} The main contribution to the cross section comes from target nuclei with 
velocity close to $v_{\rm G}$. 
When $V \leq \sqrt{\langle v_\parallel ^2 \rangle^{1/2} \; V_0}$,
this velocity is larger than the typical thermal 
velocity $\langle v_\parallel ^2 \rangle ^{1/2}$.
This result is  equivalent to the Gamow peak energy in stars, which is significantly
higher than  the thermal energy $kT$.
In terms of the energy, by putting $E_{2}=\frac{1}{2} m_2 v_G ^2$
in eq.~(\ref{E_gain}), we see that the ``most probable'' collision energy is
\footnote{The most probable energy $E_{mp}$ has not to be confused with the effective 
energy $E_{eff}$.}:
\begin{equation}
\label{Eeff}
 E_{\rm mp}= E+ 2\left(\frac{m_1}{m_1+m_2}\right)\frac{V_{0}}{V}\;E_{\rm t}\quad, 
\end{equation}
where $E_{\rm t}=\frac{1}{2} m_2 \langle v_\parallel^2 \rangle =\frac{1}{2} kT$
is the average thermal energy associated with the motion in the collisional 
direction.

{\em iii)} The energy dependence of eq. (\ref{fthermal}), 
\begin{equation} 
\label{fthermE}
f= \exp \left [\frac{1}{2} \left(\frac{m_1}{m_1+m_2} \right )\, \frac{E_t E_0}{E^2} \right ],
\end{equation}
where $E_0= \frac{1}{2} \mu V_0^2$, is different from 
that resulting from electron screening $f = \exp(D/E^{3/2})$.

{\em iv)} The resulting effects are anyhow extremely tiny. For example, for 
$d+d$ collisions ($V_0=e^2/\hbar $)  at $E=1$  keV $(V=1/5 \,V_0)$ 
and room temperature ($\langle v_\parallel ^2 \rangle^{1/2}= 5\cdot 10^{-4} V_0$) 
one has $f-1\simeq 10^{-4}$. 
A 10\% enhancement would correspond to $kT \simeq 30$ eV.

{\em v)} The same method can be extended to other motions of the target 
nuclei, provided that the velocity distribution is approximately 
Gaussian and if other interactions of the nuclei during the 
collision are neglected (sudden approximation). One has to 
replace $\langle v_\parallel ^2 \rangle  $ in eq.~(\ref{fthermal}) 
with the appropriate average 
velocity associated with the motion under investigation.  
Vibrations of the target nucleus inside a molecule or  a crystal 
lattice can be treated in this way, since the vibrational times 
are much longer than the collision times. These and other similar 
effects will be discussed in sect.~\ref{appli}.

\begin{figure}
\centerline{\hbox{
\epsfig{figure=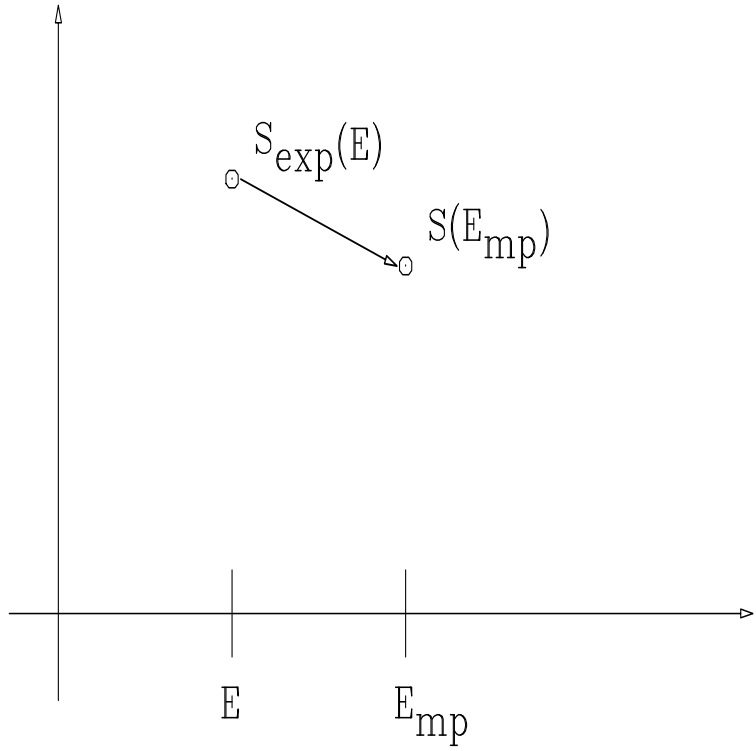,height=10cm,width=10cm,angle=90}}}
\vskip1cm
\caption[a]{Extraction of the $S$-factor from experimental data.}
\label{figEeff}
\end{figure}

{\em vi)} From the discussion presented above one gets an easy procedure to 
correct the experimental results for taking into account the 
effect of an energy spread. If the astrophysical S-factor has been 
measured at a nominal collision energy $E=\frac{1}{2} \mu V^2$, from 
$S_{\rm exp} = \sigma_{\rm exp} \, E \, \exp(V_0/V)$, then the ``true'' S-factor is 
obtained as $ S=S_{\rm exp}/ f$, where $f$ is given by eq.~(\ref{fthermal}) 
and the ``true" energy is changed from $E$ to $E_{\rm mp}$ given 
in eq.~(\ref{Eeff}) ( Fig.~\ref{figEeff}).
 In summary, the effect of the energy spread translates 
into both a cross section enhancement and an energy enhancement.

\section{General treatment}
\label{general}

In this section we shall provide a more general discussion of the energy spread effects, 
which will substantially confirm eq.~(\ref{fthermal}) and which can be applied to a 
rather large class of processes. The main assumption 
is that the projectile motion is fast in comparison with the other motions, so 
that the sudden approximation can be used.

Let us consider a projectile with velocity $V$ impinging onto a 
thin target (density $n$ and thickness $L$), where 
energy loss can be neglected.  The interaction probability  $P$ 
is the product of the interaction probability per unit time
$\dot p= n\langle\sigma v_{\rm rel}\rangle$ with the time spent 
in the target, $L/V$. The measured counting rate
$\Lambda=\epsilon I \dot{p}$, where $I$ is the beam 
current and $\epsilon$ is the detector efficiency, is thus:
\begin{equation}
\label{rate}
\Lambda = \frac{I \epsilon n L}{V} 
\langle\sigma v_{\rm rel}\rangle
\quad .
\end{equation}
As in stars, the quantity which is physically relevant is thus 
$\langle\sigma v_{\rm rel}\rangle$, where the average has to 
be taken over the target nuclei velocity distribution.

This distribution is due to the coupling with other degrees of freedom. 
Inside an atom (or a molecule, or a crystal) the nucleus is vibrating, 
its motion is altered by 
the arrival of the projectile nucleus and the calculation of the average is complicated 
in the general case. 
However, if the velocity $V$ of the impinging particle is large in comparison with the 
velocity $v$ of the target nucleus, the problem is simplified. 
The target wave function does not have time for significant evolution during the collision 
and it  can be taken as that of the initial (unperturbed) state. 
This is the main content of the sudden approximation: the velocity distribution of the 
target nuclei $\rho(v)$ can be taken as the initial one $\rho_{\rm in}(v)$ and one has
 to compute:
\begin{equation}
\label{lambda}
\Lambda= 
\frac{I \epsilon nL}{V} 
\int d^3v \; \rho_{\rm in}(v) \;
\sigma(v_{\rm rel}) v_{\rm rel}
\quad .
\end{equation}
By using eq.(\ref{sigmadivrel}), one has thus to compute:
\begin{equation}
\label{lambda2}
\Lambda= \frac{I \epsilon nL}{V}
\int d^3v \; \rho_{\rm in}(v) \; 
\left[ \frac{2 S(v_{\rm rel})}{\mu v_{\rm rel}}\;
\exp \left(-\frac{V_0}{v_{\rm rel}} \right) \right]
\quad .
\end{equation}
We recall that  $S$ is a weakly varying function of energy, 
so that it can be taken out of the integral.

%
Since  we are assuming $V^2 \gg \langle v^2 \rangle$,
we  expand the integrand
$g= \frac{1}{v_{\rm rel}}\,\exp(-V_0/v_{\rm rel})$ 
in powers of $v$ and keep the lowest order terms:
\begin{equation}
\nonumber
\Lambda = 
\frac{2S I \epsilon nL}{\mu V} \, 
\int  d^3v \;
\rho_{\rm in}(v)
\left[ g_{v=0} + v_i (\partial_i g)_{v=0} + \frac{1}{2} 
v_i v_j (\partial_i \partial_j g)_{v=0} \right] \quad.
\end{equation}
We shall consider distributions which are symmetrical for 
inversions and rotations around the collision axis $V$. In this case 
the term linear in $v$ vanishes  and the result is:
\begin{eqnarray}
\label{lambda3}
\Lambda & = & \frac{2S I \epsilon nL}{\mu V^2}\, \exp \left( -\frac{V_0}{V} \right ) \cdot\\ \nonumber
        &  &   \{ 
                1 + \frac{\langle v_\parallel^2\rangle V_0^2 }{2 \,V^4}  
              \left ( 1 - 4\frac{V}{V_0} + 2 \left ( \frac{V}{V_0} \right ) ^2 \right ) \\ \nonumber
        &   &   + \frac{ \langle  v_\perp^2\rangle V_0^2}{2 \,V^4} 
              \left[ \frac{V}{V_0} -\left ( \frac{V}{V_0} \right )^2 \right ]    \}\quad ,
\end{eqnarray}
where the index $\parallel (\perp)$  denotes the component of the velocity 
along (transverse to) the collision axis.

The term in front of the curl bracket is the counting rate 
calculated neglecting the target energy spread. So, if we define
the enhancement factor $f$ as
{\em the ratio of  the  measured
counting rate $\Lambda$ to the rate  calculated for fixed velocity $\Lambda_V$},
 we have:
\begin{equation}
\label{f_definition}
f\equiv \frac{\Lambda}{\Lambda_V} = 
V \, \exp \left ( \frac{V_0}{V} \right ) \, \int d^3v \; 
\rho_{\rm in}(v)
\left[ \frac{1}{v_{\rm rel}} \, 
\exp \left ( -\frac{V_0}{v_{\rm rel}}  \right ) \right]
\quad ,
\end{equation}
we have now:
\begin{eqnarray}
\label{fmain}
%
f & \simeq  & 1  +  \frac{\langle v_\parallel^2\rangle \, V_0^2 }{2 \,V^4}  
              \left[1 - 4\frac{V}{V_0} + 2 \left ( \frac{V}{V_0} \right ) ^2 \right] +    
\\ \nonumber  
&  & \quad +  \frac{ \langle v_\perp^2\rangle \, V_0^2}{2 \,V^4} 
              \left[ \frac{V}{V_0} -\left ( \frac{V}{V_0} \right )^2 \right ]   \quad .
\end{eqnarray}
For a one dimensional motion ($v_{\perp}=0$) it simplifies to:
\begin{equation}
\label{fpar}
f  =  1 + \frac{\langle v_\parallel^2\rangle  V_0^2 }{2 \,V^4}  
              \left[ 1 - 4\frac{V}{V_0} + 2 \left ( \frac{V}{V_0} \right ) ^2 \right] \quad .
\end{equation}
For the case of a spherically symmetrical distribution, 
$\langle v_{\parallel}^2\rangle =1/2\langle v_{\perp}^2\rangle $ one gets:
\begin{equation}
\label{fsphe}
f  =  1 +  \frac{\langle v_\parallel^2\rangle\,  V_0^2 }{2 \,V^4}  
              \left ( 1 - 2\frac{V}{V_0}  \right )\quad .
\end{equation}
This equation can be easily compared with the result of the 
previous section concerning the thermal energy effect.
By expanding eq. (\ref{fthermal}) one gets:
\begin{equation}
f = 1 + \frac{1}{2} \frac{\langle v_{\parallel}^2\rangle \, V_0^2}{V^4} \quad .
\end{equation}
This  is the same as eq.~(\ref{fsphe}) apart for the last term 
which is negligible at small velocities, since it is a higher 
order contribution in $V/V_0$.
Note that this last term arises from  the variation of the 
pre-exponential factor
$1/v_{\rm rel}$, which was neglected in the simplified 
treatment of sect.~\ref{sec2}. Clearly this term, once averaged over the 
target distribution, is smaller than $1/V$ and therefore it provides 
a reduction of the rate, as implied by the negative coefficient in 
eq.(\ref{fsphe}).

The previous results  have been obtained by neglecting higher order terms in the expansion 
of $g$. Their contribution is suppressed by a factor 
$\langle v_{\parallel}^2\rangle \, V_0^2 / V^4$.Thus 
the previous results 
are not valid for 
$V \ll \sqrt{\langle v^2 \rangle^{1/2} \; V_0}$, 
as can be simply understood.
In this case, one cannot expand the integrand function $g(v)$, since
it changes faster than the distribution function $\rho(v)$ 
over a large range of target velocities. More precisely,
the decrease of $\rho(v)$ is counter-balanced by 
the increase of $g(v)$ in a velocity range which is typically 
larger than the average target velocity dispersion 
$\langle v^2 \rangle^{1/2}$. As a consequence, the
tails of the distribution function $\rho(v)$ 
give a relevant contribution to the counting rate, 
leading to an increase of the factor $f$
with respect to the simple estimate eq.(\ref{fsphe}).

It is difficult to obtain a general expression for $f$ 
in this low velocity regime. The factor $f$ depends, in fact, 
on the shape of the distribution function. 
In the case of a gaussian distribution function, 
$\rho(v) \propto \exp(-v^{2}/2\langle v^{2} \rangle)$, 
one can use the Gamow ``trick'' described in 
the previous section which leads to eq.(\ref{fthermal}). 
For distribution functions
which decrease more slowly with $v$ one expects larger effects.

In order to have, however, a general result for the low velocity
($\langle v^2 \rangle < V^2 < \langle v^2 \rangle^{1/2} V_0$)
 behaviour of $f$, 
we note that, being the counting rate $\Lambda$ an increasing function of the
projectile velocity $V$, one has:
\begin{equation}
\label{lambdalarger}
\Lambda \ge \Lambda(0) \equiv  \frac{I \epsilon nL}{V}\;A\;
\int d^3v \; \rho_{\rm in}(v) \; 
\left[ \frac{1}{v}\;
\exp \left(-\frac{V_0}{v} \right) \right]
\end{equation} 
This means that the enhancement factor $f$ should be larger than:
\begin{equation}
\label{flarger}
f_0 =
V \exp \left(\frac{V_{0}}{V}\right) 
\int d^3v \; \rho_{\rm in}(v) \; 
\left[ \frac{1}{v}\;
\exp \left(-\frac{V_0}{v} \right) \right]  \quad .
\end{equation}

\section{Applications}
\label{appli}

The method developed in the previous sections, summarized in eq.~(\ref{fthermal})
 or in the more accurate eq.~(\ref{fmain}), can be 
applied to several  motions of the target nuclei (vibrations 
inside an atomic, molecular or crystal system), provided that 
interactions with other degrees of freedom during the collision 
can be neglected. Simply, one has to compute the value of $\langle v^2\rangle $ 
which is appropriate to the system under consideration. Also, the 
treatment can be easily extended to the effect of beam energy 
width and straggling.

\subsection{Nuclear motion inside the atom}

Very much as the motion of a star in the sky is affected by the 
presence of planets around it, the nucleus inside an atom is 
vibrating around the center of mass of the atomic system.  The 
nuclear  momentum distribution $P(p)$ is immediately determined 
from that of the atomic electrons $P_e(p_e)$ by requiring that the 
total momentum of the atom vanishes in the center of mass 
$(p=-p_e)$, where $p_e$ is the (total) momentum carried by the 
electron(s), i.e. $P(p)= P_e(-p_e)$ and the initial  nuclear 
velocity distribution  $\rho_{\rm in}(v)$ is immediately determined 
from $v=p/m_2$, where $m_2$ is the target nucleus mass.

For the case of  Hydrogen (isotope) in the ground state, the atomic electron  
momentum distribution is:
\begin{equation}
 P_e(p_e)=\frac{8}{\pi^2} \;\frac{(m_e v_0)^5}{(p_e^2 + m_e^2 v_0^2)^{4}}
\end{equation}
 so that the nucleus velocity distribution is:
\begin{equation}
\rho_{\rm in}(v)= \frac{8}{\pi^2} \; \frac{u_0^5}{(v^2 + u_0^2)^{4}} \quad ,
\end{equation}
where $u_0= (m_e/m_2) v_0 = (m_e/m_2) e^2/\hbar$, is the typical 
velocity associated with the target nuclear motion. 
In practice, this is definitely smaller than the collision 
velocity $V$, so that the sudden approximation holds and the 
results of the previous section can be applied. 

One can easily evaluate that:
\begin{equation} 
\langle v^2_\parallel \rangle = \frac{1}{3}\,u_0^2=\frac{1}{3}\left(\frac{m_e}{m_2}\right)^{2} \,v_0^2 \quad, 
\end{equation}
so that for Hydrogen-Hydrogen (or deuterium-deuterium) collisions, 
for which $V_0 = Z_1 Z_2 e^2/ \hbar = v_0$, by using   
eq.~(\ref{fsphe}), one obtains for  the enhancement factor :
\begin{equation}
\label{fat}
f_{\rm at} = 1 + \frac{1}{6}  \, \left ( \frac{m_e}{m_2} \right )^2 \, 
\left ( \frac{V_0}{V} \right )^4 \left(1-2\frac{V}{V_{0}} \right)
\quad .
\end{equation}
This is an extremely tiny correction, since one has $f_{\rm at} - 1 \simeq 2 \cdot 10^{-5}$ 
for a d-d collision at $E= 1$~keV energy .

\begin{figure}
\centerline{\hbox{
\epsfig{figure=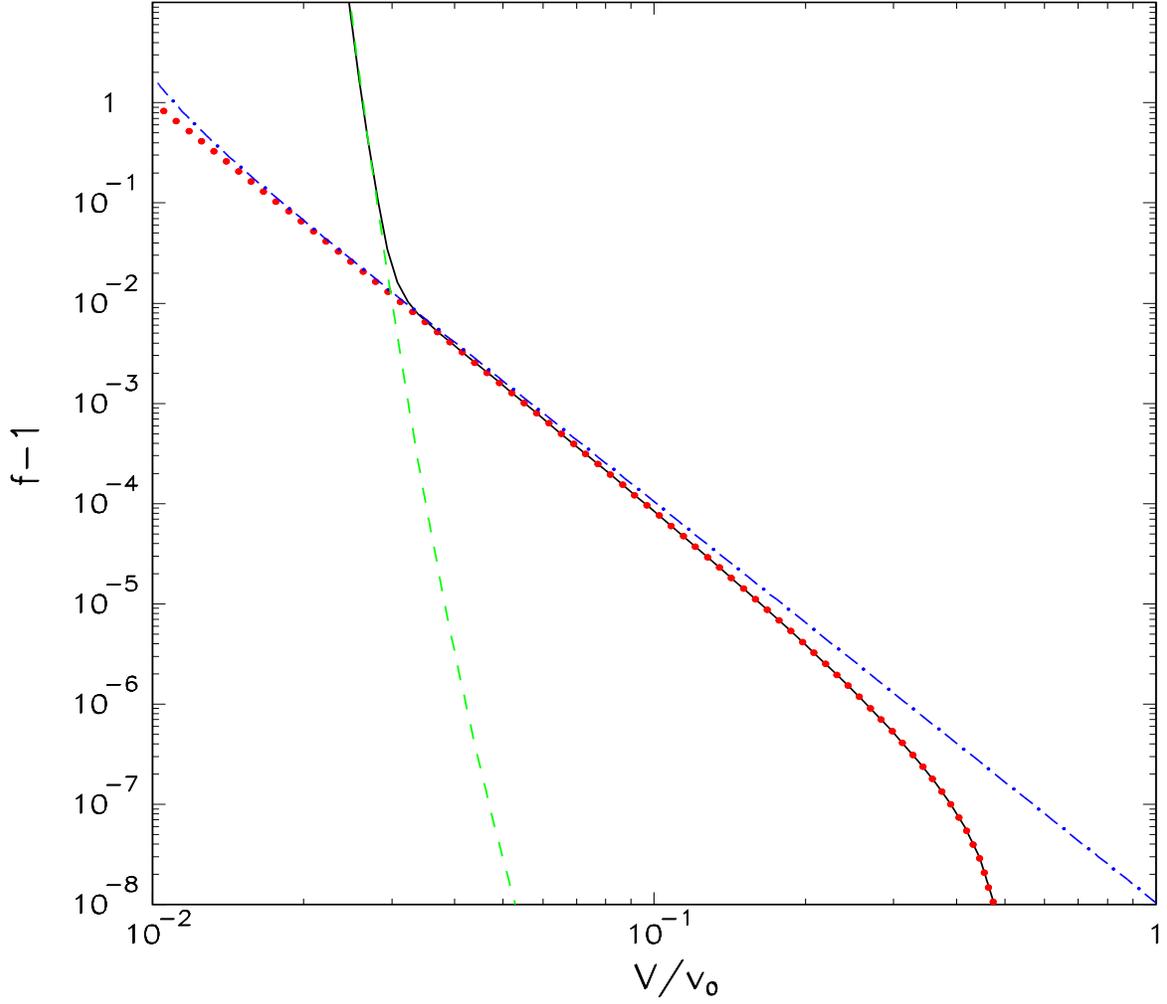,height=20cm,width=16cm,angle=0}}}
\vskip-1cm
\caption[a]{Fusion enhancement due to nuclear motion inside a H atom.
we present the numerical evaluation of eq.(\ref{fmain}) (full line), 
the approximations of  eq.(\ref{fthermal}) (dotdashed)
and of  eq.(\ref{fsphe}) (dotted), the low velocity limit eq.(\ref{ffatt}) (dashed).}
\label{figcompare}
\end{figure}

In the low energy regime, i.e. when 
$V \leq \sqrt{u_0 \; V_0} = (m_e/m_2)^{1/2} \, v_0$,
the previous estimate has to be corrected to take into account the 
contribution of the tails of the distribution function. By using eq.(\ref{flarger})
we can easily estimate:
\begin{equation}
\label{ffatt}
(f_{\rm at})_0 \simeq \frac{32\cdot 5!}{\pi}
\frac{V}{V_{0}}\left( \frac{u_0}{V_{0}}\right) ^{5}
\exp (\frac{V_{0}}{V})
\end{equation} 
In Fig. \ref{figcompare} we compare  the approximate expressions 
with the numerical evaluation of eq.~(\ref{fmain}).
In the whole range a good approximation to the full numerical calculation is provided by
$f = f_{at} + (f_{\rm at})_0$. 

\subsection{Molecular vibrations}

Let us consider, as an example, reactions involving a deuterium 
nucleus bound in a $D_2$ molecule. The target nucleus is  
vibrating, the vibration energy in the ground state being 
$E_{\rm vib}=0.19$ eV . This energy is shared between the two nuclei 
and between potential and kinetic energy, so  that the average 
kinetic energy of each nucleus is 
$\frac{1}{2} m_{d} \langle v^2\rangle _{\rm vib}=1/4 E_{\rm vib}$. 
The target nucleus  velocity,  $\langle v^2\rangle _{\rm vib}  \simeq 10^{-6}v_0^2 $, 
is much smaller than the projectile velocity so that the sudden 
approximation applies again.  By using eq.~(\ref{fthermal}) and 
assuming a random orientation of the molecular axis , 
$\langle v_\parallel^2\rangle  =1/3 \langle v^2\rangle _{\rm vib}$, we get:
\begin{equation}
f_{\rm mol}= \exp\left [\frac{\langle v^2\rangle _{\rm vib} V_0^2}{6 V^4} 
\right ] \quad .
\end{equation}
This corresponds to a  $10^{-4}$ correction at E=1 keV.
Conversely, an enhancement correction of 10\% would correspond to
$E_{vib}\simeq 200$ eV.

\subsection{Local vibrations in a crystal lattice}

When a deuterium nucleus is implanted in a crystal, it generally 
occupies an interstitial site where it performs local vibrations.
The vibration energy $E_{\rm cr}$ depends on the host lattice, being 
typically in the range of  0.1~eV, very similar to the 
molecular vibration scale. Effects associated with vibrations in 
the crystal are thus similar to those calculated for the $D_2$ 
molecule:
\begin{equation}
f_{\rm cr} \simeq f_{\rm mol} \quad .
\end{equation}

\subsection{Finite beam width and straggling}

In an ideal accelerator all projectiles have the same energy 
$E_{\rm lab}$. Actually, due to several physical processes (voltage 
fluctuations, different orbits...) the beam will have a finite 
energy width $\Delta$. As an example, in the LUNA 
accelerator one has $\Delta  \simeq 10$ eV. Furthermore, when the 
beam  passes through the target, fluctuations in the energy loss 
will produce an enlargement of the energy width (straggling).  
Thus, even neglecting the target motion, there is a collision 
energy spread.
The beam energy distribution,                                                   
\begin{equation}                                                               
P(E^{'}) \simeq \exp \left [-\frac{(E^{'}-E_{\rm lab})^2}{2\Delta^2} \right ]
\end{equation}                                                                  
gives a velocity ditribution with:  
\begin{equation}                                                                
\langle v_\parallel\rangle ^2 = 
\frac{ \Delta^2 }{  m_d E_{\rm lab}}          
\end{equation}                                                                  
                                                                                
By using eq. (\ref{fmain}) the enhancement factor is thus
\footnote{                                   
For the sake of precision, the counting rate is now:
$\Lambda=\epsilon I nL \langle \sigma\rangle _{beam}$. This is different from eq. 
(\ref{rate}). A calculation of the average, similar to that 
presented  in sect. \ref{general},  yields the same expression 
as in eq.(\ref{lambda3})
for the 
leading term in ${V/V_0}$ and different numerical coefficients for 
the higher order (negligible) terms. 
         }:                                             
\begin{equation}   
 f=\exp \left [ \frac{V_0^2 \Delta^2}{ m_1^2 V^6} \right ]   
\end{equation}

Effects are very small in the case of LUNA: for d+d at $E=1$ keV                
and $\Delta=10$ eV one has $f-1\simeq 2 \cdot 10^{-5}$.                         
The effect behaves quadratically with $\Delta$ and it can be                    
significant if momentum resolution is worse. 
Conversely, an enhancement correction of 10\% corresponds to 
$\Delta \simeq 250 $ eV.

\subsection{Polynomial velocity distributions}

One could suspect that  velocity distributions of
different shape  can provide  enhancements significantly
larger than the tiny effects which we have found so far.

In this spirit, let us consider  the case of a polynomial
velocity distribution,
\begin{equation}
\label{poly}
\rho(v)= \frac{A} {\left(v^2+B^2\right)^n}
\end{equation}
where the slowly decreasing tail should provide a
significant enhancement. Clearly the more  favourable
cases correspond to  small values of $n$.
The requirement that $\langle v^2 \rangle$ is finite implies $n\ge 3$,
so we consider $n=3$
in order to maximize the tail effect. The normalized
distribution is in this case:
\begin{equation}
\label{dist}
\rho(v)=  \frac{4}{\pi^2 \cdot 3^{3/2}}\, 
\frac{\langle v^2 \rangle^{3/2}}{\left[v^2+ (1/3) \langle v^2 \rangle \right]^3}
\end{equation}
The low energy enhancement factor $f_0$ of eq.~(\ref{flarger})
becomes now:
\begin{equation}
f_0 \simeq  \frac{16\cdot  3!}{\pi \cdot 3^{3/2}} \,\left[\langle v^2 \rangle /V_0^2\right]^{3/2}\,
\exp\left(\frac{V_0}{V}\right)  \frac{V}{V_0}
\end{equation}

In order to have $f_0 \simeq 1.1$ for d+d collisions at
$E=1$~keV one needs $\langle v^2 \rangle \simeq 3\cdot 10^{-2} V_0^2$, 
which corresponds to an average energy in the range of 1 keV,
well above the physical scale of the process.

\section{Concluding remarks}                                                    
                                                                                
We summarize the main points of this paper:\\                                     
                                                                                
\begin{itemize}                                                                
\item[\em i)]
Energy spread is a mechanism which generally provides fusion                    
enhancement.

\item[\em ii)]                                                                           
We have found a general expression for calculating the                         
enhancement factor $f$:
\begin{equation}                                                                
      f= \exp \left [ \left ( \frac{Z_1 Z_2 e^2}{\hbar} \right  )^2
           \frac{\langle v_\parallel\rangle ^2}{2 V^4} \right ]        
\end{equation}                                                                  

\item[\em iii)]                                                                           
We have provided quantitative estimates for the enhancement                     
effects. For a d+d  collision one has:

\begin{tabular}{lcl} 
thermal motion:     &  & $f-1 \simeq 10^{-4}$ (E/1keV)$^{-2}$ \\ 
vibrational motion: &  & $f-1 \simeq (10^{-5}$ -- $10^{-4}$) (E/1keV)$^{-2}$ \\ 
beam width:         &  & $f-1 \simeq 10^{-5}$ (E/1keV)$^{-3}$ \\ 
\end{tabular}

\item[\em iv)]                                                                           
All these effects are marginal at the energies which are presently              
measurable, however they have to be considered  in                    
future experiments at still lower energies.

\item[\em v)]                                                                           
This study allows to exclude several effects as possible                        
explanation of the observed anomalous fusion enhancements,                      
which remain a mistery.
\end{itemize}




\begin{table}
\caption[a]{Summary of  results for electron
screening effects}
\begin{tabular}{llll}
Reaction & $U_{ex}$ (eV) &  $U_{ad}$$\,\,^a$ (eV) & Ref.\\
\hline
          d(d,p)t       & $ 25\pm  5$$\,\,^b$ & 28.5 & \cite{Greife}\\
$^3$He(d,p)$^4$He       & $219\pm  7$ & 114 & \cite{Aliotta}\\
d($^3$He,p)$^4$He       & $109\pm  9$ & 102 & \cite{Aliotta}\\
$^3$He($^3$He,2p)$^4$He & $294\pm 47$ & 240 & \cite{Luna2}\\
$^3$He($^3$He,2p)$^4$He & $432\pm 29$ & 240 & \cite{Junker}\\
$^6$Li(p,$\alpha$)$^3$He & $470\pm 150$ & 184 & \cite{Engstler}\\
$^6$Li(d,$\alpha$)$^4$He & $320\pm 50$ & 184 & \cite{Musumarra}\\
$^7$Li(p,$\alpha$)$^4$He & $330\pm 40$ & 184 & \cite{Lattuada}\\
$^9$Be(p,d)$^8$Be & $900\pm 50$ & 262 & \cite{Zahnow}\\
$^{11}$B(p,$\alpha$)$^8$Be & $430\pm 90$ & 346 & \cite{Angulo}\\
\end{tabular}
{\footnotesize 
 $^a$ Values calculated for atomic target, following
\cite{Bracci90}. It is assumed that at fusion  hydrogen projectiles
are charged or neutral with equal probability. Helium projectiles are 
assumed to be $He^+ (He)$ with 20\% (80\%) probability.\\
 $^b$ This value results from gaseous target. Much larger values have been found
when deuterium is implanted in metals \cite{deuterium}.
} 
\label{Tabrolf}
\end{table}


\end{document}